\documentclass[aps,pra,floatfix,twocolumn,showpacs]{revtex4}
\usepackage{epsfig}
\usepackage{graphicx}
\usepackage{dcolumn}
\begin{document}
\title{\bf Calculations of energy levels and lifetimes of low-lying states 
of barium and radium}
\author{V. A. Dzuba}
\email{V.Dzuba@unsw.edu.au}
\affiliation{School of Physics, University of New South Wales, Sydney 2052, Australia}
\author{J. S. M. Ginges}
\email{ginges@phys.unsw.edu.au}
\affiliation{School of Physics, University of New South Wales, Sydney 2052, Australia}
\affiliation{Department of Physics, University of Alberta, Edmonton AB T6G 2J1, Canada}

\date{\today}

\begin{abstract}

We use the configuration interaction method and many-body perturbation theory
to perform accurate calculations of energy levels, transition amplitudes, 
and lifetimes of low-lying states of barium and radium. Calculations for
radium are needed for the planning of measurements of parity and
time invariance violating effects which are strongly enhanced in this
atom. Calculations for barium are used to control the accuracy of the 
calculations.

\end{abstract}

\maketitle

\section{introduction}

Measurements of the effects of parity ($P$) and time ($T$) invariance violation in atoms 
are an effective means to search for new physics beyond the standard model 
(see, e.g., Ref. \cite{Ginges}). The radium atom is a good candidate for this 
search due to the sizeable enhancement of the $P$- and $T$-odd effects 
arising from the nuclear and electronic structures \cite{FlambaumRa,DzubaRa}.
Preparations for measurements are currently in progress 
at Argonne National Laboratory \cite{argonne} and KVI \cite{groningen}. 

In our previous work \cite{DzubaRa} we performed a detailed study of those
$P$- and $T$-odd effects in radium that are enhanced due to the close proximity of 
states of opposite parity. Estimates of the lifetimes of all low states of radium 
were also presented. 

A detailed knowledge of the energy levels and transition amplitudes of
radium is extremely important at the first stages of the experiment
when the trapping and cooling scheme is developed and tested. Energy
levels of radium presented in Moore's book \cite{Moore} are based
on works by Rasmussen \cite{Rasmussen} and Russell \cite{Russell}
conducted as early as 1934. The first work \cite{Rasmussen} 
presents measurements of transition frequencies 
while the second work \cite{Russell} corrects the interpretation
of these measurements. To the best of our knowledge no further
measurements were performed for radium since that time.
There is some concern ignited by recent calculations by 
Biero\'{n} {\em et al.} \cite{F-F} that the positions of the
energy levels of the $7s6d$ configuration of radium might be
lower then those presented in Moore's book~\cite{Moore}. 
This, if true, can totally destroy the cooling scheme 
adopted by the Argonne group \cite{argonne}. 

The actual position of the $7s6d~{^3D}_2$ energy level
is also important for the enhancement of the $P$- and $T$-odd
effects considered in Refs. \cite{FlambaumRa,DzubaRa}. 
This enhancement is due to the very small energy interval
($\sim 5~{\rm cm}^{-1}$) between states of opposite parity
$7s7p~{^3P}_1$ and  $7s6d~{^3D}_2$. Any significant change 
in the position of either state would also destroy the
enhancement.

We haven't performed accurate calculations of radium energy levels before.
However, calculations for barium~\cite{Johnson98,vn}, which has a similar 
electron structure, show that such calculations are possible.
The theoretical uncertainty cannot be as low as 5~cm$^{-1}$
needed to confirm the strong $P$-odd enhancement due to the small energy interval
between states of opposite parity. However, it can be small enough to
address any concern about the experimental numbers.

In the present work we perform accurate calculations of the energy levels
and $E1$ and $E2$ transition amplitudes for low states of radium and
barium. We use the $V^{N-2}$ approximation (see, e.g., Ref. \cite{vn}). 
Relativistic Hartree-Fock calculations are carried out for 
a doubly ionized ion with both valence electrons removed. The self-consistent
potential of the core (the $V^{N-2}$ potential) is used to construct
the effective Hamiltonian for the configuration interaction (CI) treatment
of the valence electrons. Core-valence correlations are also included by
means of many-body perturbation theory (MBPT). Certain classes of
dominating higher-order diagrams for the core-valence correlation are 
included in all orders in the Coulomb interaction.

The barium and radium atoms have similar electron structure, but more experimental data is available
for barium. Parallel calculations for these atoms provide a 
control of the accuracy. The resulting accuracy for the
energies of barium and radium is a fraction of a percent or better
for removal energies and for the intervals between ground and low-lying states. 
There is also very
good agreement between experimental and calculated lifetimes of
several states of barium. This means that the accuracy of the results
for radium should also be very high.

In the end we see no reason for not trusting the experimental energies
of radium presented in Moore's book. Therefore, the trapping and cooling 
of radium should work as planned.

\section{Calculations}

We use the combined configuration interaction and many-body perturbation 
theory method (CI+MBPT, \cite{Kozlov96}) and the $V^{N-2}$ approximation (see \cite{vn}) 
to perform the calculations. Like in the standard CI method, the Schr\"{o}dinger
equation for the wave function of two valence electrons is written in matrix form
\begin{equation}
  (\hat H^{\rm eff} - E)\Psi = 0.
\label{Schr}
\end{equation}
$\Psi$ is expressed as an expansion over single-determinant two-electron
wave functions
\begin{equation}
  \Psi = \sum_i c_i \Phi_i(r_1,r_2).
\label{psi}
\end{equation}
$\Phi_i$ are constructed from the single-electron valence basis 
states calculated in the $V^{N-2}$ potential. 
$E$ in Eq. (\ref{Schr}) is the valence removal energy 
(energy needed to remove two valence electrons from the atom).

The effective Hamiltonian has the form
\begin{equation}
  \hat H^{\rm eff} = \hat h_1(r_1) + \hat h_1(r_2) + \hat h_2(r_1,r_2),
\label{heff}
\end{equation}
$\hat h_1(r_i)$ is the one-electron part of the Hamiltonian
\begin{equation}
  \hat h_1 = c \mbox{\boldmath$\alpha$}\cdot {\bf p} + (\beta -1)mc^2 - \frac{Ze^2}{r} + V^{N-2}
 + \hat \Sigma_1.
\label{h1}
\end{equation}
$\hat \Sigma_1$ is the correlation potential which represents the correlation
interaction of a valence electron with the core. It is essentially
the same as for atoms with one valence electron (see, e.g., \cite{CPM,Dzuba89,vn}).

$\hat h_2$ is the two-electron part of the Hamiltonian
\begin{equation}
  \hat h_2 = \frac{e^2}{|\mathbf{r_1 - r_2}|} + \hat \Sigma_2(r_1,r_2),
\label{h2}
\end{equation}
$\hat \Sigma_2$ is the two-electron part of core-valence correlations. It represents 
screening of the Coulomb interaction between valence electrons by core electrons.

The terms $\hat \Sigma_1$ and $\hat \Sigma_2$ can be calculated using 
MBPT. The expansion starts from the second order and inclusion of the
second-order core-valence correlations into the effective CI
Hamiltonian is very important for obtaining good agreement with
experiment (see, e.g., \cite{Kozlov96,Johnson98}). However, as demonstrated in 
Ref. \cite{vn}, inclusion of higher-order core-valence
correlations leads to further significant improvement of the results.

It is convenient to start the calculations from a positive ion (Ba$^+$
or Ra$^+$) with one valence electron. The equation for
a single-electron valence state $v$ has the form
 \begin{equation}
  (\hat h_1 +\hat \Sigma_1 - \epsilon_v)\psi_v = 0,
\label{Brueck}
\end{equation}
where $\hat h_1$ is given by Eq. (\ref{h1}). Solving this equation produces
so-called Brueckner orbitals $\psi_v$ and energies
$\epsilon_v$ in which correlations with the core are included by means
of the correlation potential $\hat \Sigma_1$. Comparing $\epsilon_v$
with the experimental spectrum of the positive ion is a way to study
different approximations for $\hat \Sigma_1$. Table \ref{ions} compares 
energy levels of Ba$^+$ and Ra$^+$, calculated in
different approximations, with experiment.
Hartree-Fock (HF) energies correspond to $\hat \Sigma_1 = 0$ in Eq. (\ref{Brueck}).
$\hat \Sigma^{(2)}$ is the correlation potential calculated in the
second order of MBPT.  $\hat \Sigma^{(\infty)}$ is the correlation
potential in which two classes of higher-order diagrams are
included in all orders. These are the {\em screening of the Coulomb interaction}
and the {\em hole-particle interaction}. This is done in exactly the same way
as in our previous works for atoms with one valence electron
(see, e.g., \cite{Dzuba89,DzubaFr,DzubaCs,Ginges}). One can see
from the table that inclusion of core-valence correlations
is very important for obtaining accurate results and inclusion 
of higher-order core-valence correlations leads to further
significant improvement. The energies obtained
with $\hat \Sigma^{(\infty)}$ are within 0.2-0.3\% of the experimental values 
for $s$ and $p$ states, and less accurate for $d$ states.

The column marked $\Delta_c$ in Table \ref{ions} lists
correlation energies (the difference between Hartree-Fock and experimental energies) 
of valence states of Ba$^+$ and Ra$^+$.
One can see that the largest correlation energy is in $d$ states.
This means that these states are more sensitive to the treatment
of the correlations and generally are harder to calculate to high
accuracy. This is why the accuracy for $d$ states of Ba$^+$ and
Ra$^+$ is not as good as for $s$ and $p$ states. Large core-valence
correlations for $d$ states also manifest themselves in the
energies of two-electron configurations containing $d$-electrons
(e.g., $6s5d$ configurations of Ba and $7s6d$ configurations of
Ra, see discussion below). 

\begin{table}
\caption{\label{ions}Energy levels of Ba$^+$ and Ra$^+$ in different
approximations. Energies are given in cm$^{-1}$ with respect to the continuum, 
minus sign is omitted. $\Delta_c = $~E(exp)~$-$~E(HF),
$\Delta = $~E(exp)~$-$~E($\hat \Sigma^{(\infty)}$).}
\begin{ruledtabular}
\begin{tabular}{ccccccc}
State &   Exp.\cite{Moore} &  HF    & $\Delta_c$ & $\hat \Sigma^{(2)}$ 
                               & $\hat \Sigma^{(\infty)}$ 
                                        &  $\Delta$ \\
\hline
\multicolumn{6}{c}{Barium} \\
$6s_{1/2}$ &  80687 & 75339 & 5348  &  82318  &  80816  & -129 \\
$5d_{3/2}$ &  75813 & 68139 & 7674  &  77224  &  76345  & -532 \\
$5d_{5/2}$ &  75012 & 67665 & 7347  &  76286  &  75507  & -495 \\
$6p_{1/2}$ &  60425 & 57265 & 3160  &  61180  &  60603  & -178 \\
$6p_{3/2}$ &  58734 & 55873 & 2861  &  59388  &  58879  & -145 \\
\multicolumn{6}{c}{Radium} \\
$7s_{1/2}$ &  81842 & 75898 & 5944  &  83826  &  81988  & -146 \\
$6d_{3/2}$ &  69758 & 62356 & 7402  &  71123  &  70099  & -341 \\
$6d_{5/2}$ &  68099 & 61592 & 6507  &  69101  &  68392  & -293 \\
$7p_{1/2}$ &  60491 & 56878 & 3613  &  61386  &  60702  & -211 \\
$7p_{3/2}$ &  55633 & 52906 & 2727  &  56245  &  55753  & -120 \\
\end{tabular}
\end{ruledtabular}
\end{table}

Calculations for positive ions give us a very good approximation for 
the $\hat \Sigma_1$ operator in the Hamiltonian (\ref{heff}) for a 
two-electron system. However, we also need to calculate the two-electron
operator $\hat \Sigma_2$.
We calculate it in the second-order of MBPT. Formally, the MBPT expansion for $\hat \Sigma_1$ and
$\hat \Sigma_2$ goes over the same orders of perturbation
theory. However, numerical results show that an accurate treatment
of $\hat \Sigma_1$ is usually more important than that of $\hat \Sigma_2$.
Although inclusion of the higher-order correlations into $\hat \Sigma_2$
may lead to further improvement of the results, we leave this at the
moment for future work.

\subsection{Energies of barium and radium}

\begin{table*}
\caption{\label{bara}Ground state ($^1S_0$) removal energies (a.u.) and excitation 
energies (cm$^{-1}$) of low states of barium and radium.
$\Delta_c = $~E(exp)~$-$~E(CI),
$\Delta = $~E(exp)~$-$~E($f_2\hat \Sigma_1^{(\infty)}$).}
\begin{ruledtabular}
\begin{tabular}{ll rrrr rrrrr}
\multicolumn{2}{c}{State}  & \multicolumn{1}{c}{Exp.~\cite{Moore}} & 
\multicolumn{1}{c}{CI} & \multicolumn{1}{c}{$\Delta_c$} & 
\multicolumn{1}{c}{$\hat \Sigma^{(2)}$} & \multicolumn{1}{c}{$\hat \Sigma^{(\infty)}$} & 
\multicolumn{1}{c}{$f_1\hat \Sigma_1^{(\infty)}$} &
\multicolumn{1}{c}{$f_2\hat \Sigma_1^{(\infty)}$} & \multicolumn{1}{c}{$\Delta$} &
Other  \\
\hline

\multicolumn{10}{c}{Barium} \\

$6s^2$ & $^1S_0$ & -0.55915 &   -0.52358 &       &   -0.56996 &  -0.55903 & -0.55799 & -0.55915  & \\
$6s5d$ & $^3D_1$ &     9034 &      11585 & -2551 &       8956 &      8425 &  8730 & 9040 & -6 & \\
       & $^3D_2$ &     9216 &      11662 & -2446 &       9165 &      8611 &  8910 & 9217 & -1 & \\
       & $^3D_3$ &     9597 &      11835 & -2228 &       9609 &      8999 &  9283 & 9582 & 14 & \\
  		        	    		 		    	 
       & $^1D_2$ &    11395 &      12833 & -1438 &      11733 &     11020 & 11323 & 11627 & -232 & \\
		  				 			     
$6s6p$ & $^3P_0$ &    12266 &       9947 &  2319 &      13088 &     12377 & 12400 & 12270 & -4 & \\
       & $^3P_1$ &    12637 &      10278 &  2359 &      13466 &     12740 & 12754 & 12638 & -1 & \\
       & $^3P_2$ &    13515 &      11019 &  2496 &      14374 &     13611 & 13596 & 13518 & -3 & \\
		        	    		 		    	 
       & $^1P_1$ &    18060 &      16919 &  1141 &      18631 &     17778 & 17832 & 17834 & 227 & \\ 
		        	    		 		    	 
$5d6p$ & $^3F_2$ &    22065 &      22018 &    47 &      22765 &     21502 & 21828 & 22041 & 23 & \\
       & $^3F_3$ &    22947 &      22573 &   376 &      23732 &     22403 & 22698 & 22926 & 21 & \\
       & $^3F_4$ &    23757 &      23172 &   585 &      24624 &     23230 & 23500 & 23746 & 11 & \\

\multicolumn{10}{c}{Radium} \\

$7s^2$ & $^1S_0$ & -0.56690&-0.52546 &&-0.58071 &-0.56687&-0.56567&-0.56695 &&-0.57979\footnotemark[1] \\

$7s7p$ & $^3P_0$ & 13078 & 10380 &  2698 & 14202 & 13277 & 13293 & 13132 & -53 & 14268\footnotemark[1] \\
       & $^3P_1$ & 13999 & 11240 &  2759 & 15118 & 14161 & 14166 & 14027 & -28 & 15159\footnotemark[1] \\
       & $^3P_2$ & 16689 & 13473 &  3216 & 17879 & 16813 & 16764 & 16711 & -22 & 17937\footnotemark[1] \\
$7s6d$ & $^3D_1$ & 13716 & 15231 & -1515 & 14043 & 13342 & 13423 & 13727 & -11 & 14012\footnotemark[1] \\ 
       & $^3D_2$ & 13994 & 15284 & -1290 & 14371 & 13612 & 13683 & 13980 &  14 & 14465\footnotemark[1] \\
       &         &       &       &       &       &       &       &       &     & 12958\footnotemark[2] \\ 
       & $^3D_3$ & 14707 & 15461 &  -754 & 15254 & 14323 & 14364 & 14642 & 65 & 15921\footnotemark[1] \\
       & $^1D_2$ & 17081 & 16798 &   283 & 18052 & 17007 & 17060 & 17333 & -252 &    \\
$7s7p$ & $^1P_1$ & 20716 & 18686 &  2030 & 21547 & 20487 & 20459 & 20450 & 266 & 21663\footnotemark[1] \\
       &         &       &       &       &       &       &       &       &     & 20835\footnotemark[2] \\
$7s8s$ & $^2S_1$ & 26754 & 24030 &  2724 & 27643 & 26673 & 26571 & 26669 & 85 &     \\
$6d7p$ & $^3F_2$ & 28038 & 26328 &  1710 & 29425 & 27736 & 27833 & 28001 & 37 &     \\
       & $^3F_3$ & 30118 & 27713 &  2405 & 31745 & 29848 & 29891 & 30077 & 41 &     \\ 
       & $^3F_4$ & 32368 & 29383 &  2985 & 34195 & 32134 & 32129 & 32370 & -2 &     \\
\end{tabular}
\end{ruledtabular}
\noindent \footnotetext[1]{Dzuba {\em et al.} \cite{KozlovRa}.}
\noindent \footnotetext[2]{Biero\'{n} {\em et al.} \cite{F-F}.}
\end{table*}

Theoretical and experimental energies of neutral barium and radium
are presented in Table \ref{bara}. Experimental values are taken from
Moore's tables \cite{Moore}. We present two-electron removal energies 
(in a.u.) for the ground state ($^1S_0$) of both atoms. The experimental
value is the sum of the ionization potential of the neutral atom
and its positive ion. Energies of excited states are given in cm$^{-1}$
with respect to the ground state. 

The CI column in Table \ref{bara} corresponds to the standard configuration
interaction method [$\hat \Sigma_{1,2}=0$ in Eqs. (\ref{h1}) and (\ref{h2})]. 
It takes into account correlations between valence electrons but neglects 
correlations between core and valence electrons. 
We use B-splines to construct the basis of single-electron  
states. 50 B-splines are calculated in a cavity of radius $40a_B$,
where $a_B$ is the Bohr radius. Eigenstates of the Hartree-Fock Hamiltonian
are constructed from these B-splines and the 14 lowest states above the
core in each of the $s$, $p_{1/2}$, $p_{3/2}$, $d_{3/2}$, $d_{5/2}$,
 $f_{5/2}$, and $f_{7/2}$ waves are used in CI calculations. The
uncertainty due to incompleteness of the basis is very low. It is
$\leq 10 {\rm cm}^{-1}$ for $s^2$ and $sp$ configurations and
$\leq 50 {\rm cm}^{-1}$ for $sd$ configurations.

The next column ($\Delta_c$) lists the difference between experimental
and CI energies. This difference is mostly due to core-valence
correlations. There are also contributions to $\Delta_c$ due
to the Breit interaction, radiative corrections, incompleteness
of the basis for valence-valence correlations, etc. However, all these
contributions are small.

We include core-valence correlations by introducing operators
$\hat \Sigma_1$ and $\hat \Sigma_2$ into the effective CI Hamiltonian
[see Eqs. (\ref{h1}) and (\ref{h2})]. 
Comparison of the $\Delta_c$ values for neutral Ba and Ra presented 
in Table \ref{bara} with the correlation energies ($\Delta_c$) for 
positive ions (Table \ref{ions}) reveals that core-valence
correlations have a larger effect on the energies of positive ions than
on neutral atoms. This is due to a cancellation of contributions
from $\hat \Sigma_1$ and $\hat \Sigma_2$. The Hamiltonian for
a positive ion (\ref{Brueck}) has only $\hat \Sigma_1$, while the 
Hamiltonian for a neutral atom (\ref{heff},\ref{h1},\ref{h2}) has both.
On the other hand, in the $V^{N-2}$ approximation used in the present work,
the $\hat \Sigma_1$ operator for a neutral atom is the same
as for a positive ion.

The cancellation between the two types of core-valence correlations
($\hat \Sigma_1$ and $\hat \Sigma_2$) has an effect on the accuracy 
of calculations. The accuracy is poorer when the cancellation is stronger.
It is easy to see that the strongest cancellation takes place for
$sd$ configurations of barium and radium. Indeed, correlation 
corrections to the energies of $d$ states of Ba$^+$ and Ra$^+$
are about two times larger than those for $s$ and $p$ states (see Table \ref{ions}).
However, corrections to the energies of $sd$ configurations of
neutral barium and radium are about the same or even smaller than
for $s^2$ and $sp$ configurations. 

The column in Table \ref{bara} marked by $\hat \Sigma^{(2)}$ lists results 
obtained with both $\hat \Sigma_1$ and $\hat \Sigma_2$ calculated
in the second order of MBPT. We use the same B-splines to calculate
$\hat \Sigma$ as for the CI calculations. However, we use 45 out of 50 
eigenfunctions and go up to $l=5$ in the partial wave expansion. 
Inclusion of second-order $\hat \Sigma$ leads to significant
improvement of the results. The remaining deviation from experiment
is just a small fraction of the total core-valence correction
$\Delta_c$. However, we do further steps in trying to 
improve the results. We replace the second-order $\hat \Sigma_1$
with the all-order operator $\hat \Sigma_1^{(\infty)}$. 
We use the Feynman diagram technique as described in our earlier 
papers \cite{Dzuba89,DzubaFr,DzubaCs} to calculate $\hat \Sigma_1^{(\infty)}$. 
The results are presented in column $\hat \Sigma_1^{(\infty)}$ of Table \ref{bara}. 
As one can see, inclusion of higher-order correlations 
into $\hat \Sigma_1$ leads to significant improvement of the removal 
energies but not of the energy intervals (see also Ref. \cite{vn}).
There are at least two reasons for this. First, the change in
$\hat \Sigma_1$ operator between the positive ion and neutral atom, 
and second, higher orders in $\hat \Sigma_2$. $\hat \Sigma$ is
an energy-dependent operator: $\hat \Sigma \equiv \hat \Sigma(\epsilon)$. 
It should be calculated at the energy of the state for which it is to be 
used. For example, $\hat \Sigma_s$ for the $6s$ state of Ba$^+$ should be 
calculated at $\epsilon = \epsilon(6s)$, etc. Using exactly the same
$\hat \Sigma_1$ operator for the positive ion and neutral atom corresponds
to an approximation in which the energy parameter for $\hat \Sigma$ is chosen
assuming that two-electron energy of a neutral atom is equal to the sum
of the two single-electron energies of a positive ion. This approximation is too 
rough and some adjustment in the energy parameter is needed.
An accurate adjustment is ambiguous. For example, the $s$-wave $\hat \Sigma$
for the $s^2$ and $sp$ configurations are not the same since these
configurations have different energies. Moreover, $\hat \Sigma$
operators for different states of the same configuration are not 
the same since different states have different energies.

In the present paper we use a simpler way of adjusting the value
of the $\hat \Sigma$ operator, leaving an accurate treatment
of its energy dependence for future work. We scale the
single-electron part of the operator $\hat \Sigma_1$ while
leaving the two-electron part $\hat \Sigma_2$ unchanged.
Numerous tests show that any reasonable change in $\hat \Sigma_2$
does not lead to a significant change in the spectra of Ba or Ra.
Therefore, we scale $\hat \Sigma_1$ to fit known energies
of Ba, Ba$^+$, and Ra$^+$ and use this scaling to calculate
energies of Ra.

The most straightforward way to scale the energy levels of Ra 
would be to perform an accurate fitting of the energy levels of
Ba and use the same scaling parameters to do calculations for
Ra. However, this method does not take into account the real 
difference in electron structure of the atoms. The ordering of the 
energy levels of Ba and Ra are different. States of the $sd$ configuration
lie below the $sp$ configuration for Ba and above the $sp$
configuration for Ra. Actually, there are more similarities between the 
neutral atom and its positive ion than between neutral Ba and Ra. 
We can use these similarities to construct a fitting procedure which 
takes into account the difference between Ba and Ra.

First, we scale $\hat \Sigma_1$ to fit the energies of
Ba$^+$ and Ra$^+$. Fitting coefficients are presented in
Table \ref{fitting}. They are slightly different for $d$-states
of Ba$^+$ and Ra$^+$ (0.94 for $d$-states of Ba$^+$ and 0.96
$d$-states of Ra$^+$). This is because the $5d$ states of Ba$^+$
are closer to the core and the correlation correction is larger.
Our $\hat \Sigma^{(\infty)}$ operator is less accurate for 
$d$ states than for $s$ and $p$ states and the larger correlation
correction leads to a noticeable loss in accuracy. However,
with the value of $\hat \Sigma_1$ reduced by only six or less
percent, energy levels of Ba$^+$ and Ra$^+$ are fitted exactly.  

Then we use the same scaling of $\hat \Sigma_1$ to calculate
the energy levels of neutral Ba and Ra. The results are presented 
in Table \ref{bara} under the $f_1\hat \Sigma_1^{(\infty)}$ mark.
There are two important things to note. The first is the 
significant improvement in the agreement with experiment. The second is
the remarkable similarity between Ba and Ra which was never that
good for any other approximation used so far. Now all $^3D$
states of both atoms are about 300~cm$^{-1}$ below the experimental
values while all $^3P$ states are about 100-200~cm$^{-1}$ above the 
experimental values. This is enough to indicate that
the experimental energies of Ra are correct or that at least
there is no reason to believe otherwise.

However, we do one more step. We change the scaling parameters 
by fitting to the energy levels of neutral barium. The new values are 
presented in Table \ref{fitting}. This change in the scaling parameters accounts
for the energy dependence of the $\hat \Sigma_1$ operator discussed 
above. Then the scaling parameters for Ra are calculated using the formula

\[ f_i({\rm Ra}) = f_i({\rm Ba})\frac{f_i({\rm Ra^+})}{f_i({\rm Ba^+})}. \]

In other words, barium scaling parameters are corrected using the
difference in fitting for Ba$^+$ and Ra$^+$. The new fitting 
parameters for Ra are also presented in Table \ref{fitting}.

The results of calculations for Ba and Ra with the new fitting
parameters are presented in Table \ref{bara} under the
$f_2\hat \Sigma_1^{(\infty)}$ mark. One can see that the
$^1S$, $^3D$, and $^3P$ states of Ba are fitted almost exactly.
The $^1D$ and $^1P$ states are less accurate because strictly
speaking one cannot use the same $\hat \Sigma$ for $^3P$ and
$^1P$ states and for $^3D$ and $^1D$ states due to the difference
in energies of these states.

Calculations for Ra with the new scaling parameters reduce the
deviation of the theoretical values from experiment to about 
50~cm$^{-1}$ or less for all low states. Let us stress once more
that no knowledge of the Ra spectrum was used to do the fitting.
Values of the scaling parameters were found by fitting the spectra
of Ba, Ba$^+$, and Ra$^+$. The good agreement of the final numbers
with experiment leaves no room for any claim that the experimental
values might be incorrect. As discussed by Russell \cite{Russell}, 
the difference between two possible ways of interpretation of the 
experimental data is 627.66~cm$^{-1}$. This is much larger than the 
difference between our calculated energies and the experimental energies of Ra. 

\begin{table}
\caption{\label{fitting}Fitting factors $f$ for rescaling of 
$\hat \Sigma^{(\infty)}$ to reproduce experimental energies of
Ba$^+$ and Ra$^+$}
\begin{ruledtabular}
\begin{tabular}{llllll}
Atom & $s_{1/2}$ & $p_{1/2}$ & $p_{3/2}$ & $d_{3/2}$ & $d_{5/2}$ \\
\hline
Ba~II & 0.9777 & 0.95  & 0.96  & 0.94   & 0.94 \\
Ba~I  & 1.0032 & 1.046 & 1.046 & 0.9164 & 0.9164 \\
Ra~II & 0.9777 & 0.95  & 0.96  & 0.96   & 0.96 \\
Ra~I  & 1.0032 & 1.046 & 1.046 & 0.9359 & 0.9359 \\
\end{tabular}
\end{ruledtabular}
\end{table}

In the last column of Table \ref{bara} we present the results of 
our previous calculations of energy levels of Ra \cite{KozlovRa} together 
with the results of recent calculations for Ra by Biero\'{n} 
{\em et al.} \cite{F-F}. Our previous calculations were very similar
to those presented in the table in the $\hat \Sigma^{(2)}$ column.
They were also obtained in the $V^{N-2}$ approximations with
the second-order $\hat \Sigma$. However, the basis of single-electron
states was different. The difference between the present $\hat \Sigma^{(2)}$
results and the results of Ref. \cite{KozlovRa} can serve as an upper
limit on the uncertainty due to incompleteness of the single-electron basis 
for the valence states. 
The real uncertainty of the present calculations is several times smaller
due to better saturation of the basis.

Calculations of Ref. \cite{F-F} were performed 
by means of the multi-configuration Dirac-Hartree-Fock method.
The authors use the results to claim that experimental energies 
of the $7s6d$ configuration of Ra might be incorrect. 
Indeed, their calculated value for the $^3D_2$ state is 1034~cm$^{-1}$ 
below the experimental value. On the other hand, the deviation from 
experiment of the only other calculated energy level, the energy of the $^1P_1$ state, 
is only 119~cm$^{-1}$. No other energy levels of Ra were calculated and no 
calculations for other two-electron systems were used to control the accuracy.
Therefore, it is hard to make any judgement about the quality of
these calculations. However, let us remind the reader that
calculations for $sd$ configurations are more difficult than 
for $sp$ configurations due to the larger correlation interaction
of the $d$ electron with the core and the stronger cancellation 
between $\hat \Sigma_1$ and $\hat \Sigma_2$ terms (see discussion
above). Therefore, the accuracy obtained for the $^1P_1$ state cannot
serve as a guide for the accuracy for the $^3D_2$ state.
Apart from that, good agreement with experiment for just one
number cannot rule out a fortunate coincidence. 

\subsection{Transition amplitudes}

The leading contribution to the amplitude of a transition between
states $v$ and $w$ of Ba or Ra is given by
\begin{equation}
  A_{vw} = \langle \Psi_w | \hat f | \Psi_v \rangle,
\label{me0}
\end{equation}
where $\Psi_w$ and $\Psi_v$ are the solutions of Eq.~(\ref{Schr})
and $\hat f$ is the operator of the external field. This expression doesn't
take into account the effect of the external field on the atomic core.
This effect, which is known as {\em core polarization}, is very
important and can change the amplitude significantly.
It can be included by means of the time-dependent Hartree-Fock
method (TDHF) which is equivalent to the well known random-phase
approximation (RPA) method.

Every single-electron core function is presented in the RPA
approximation as $\psi_a + \delta \psi_a$, where $\psi_a$ is the 
Hartree-Fock wave function of the core state $a$ calculated in 
the $V^{N-2}$ potential; $\delta \psi_a$ is the correction due
to the external field. The corrections to all core states are
found self-consistently by solving Hartree-Fock-like
equations
\begin{equation}
  (\hat H_0 - \epsilon_a)\delta \psi_a = - \hat f \psi_a - \delta V_{core} \psi_a,
\label{rpa}
\end{equation}
where $H_0$ is the Hartree-Fock Hamiltonian, $\hat f$ is the operator of the 
external field, and $\delta V_{core}$ is the correction to the self-consistent
potential of the core due to the effect of the external field. Note that in our case
$V_{core} \equiv V^{N-2}$. The $\delta V_{core}$
term is calculated using the $\delta \psi$ corrections to all core states.
The final expression for the transition amplitude has the form
\begin{equation}
  A_{vw} = \langle \Psi_w | \hat f + \delta V_{core}| \Psi_v \rangle.
\label{me}
\end{equation}
Amplitudes of electric dipole transitions (E1) between low states
of barium and radium calculated in different approximations are
presented in Table \ref{BaRaE1}. Core polarization is included
everywhere since it is known to be an important effect. We study only
the effect of core-valence correlations on the amplitude. As with energies, 
inclusion of core-valence correlations have a significant effect on the amplitudes. 
On the other hand, amplitudes calculated with
$\hat \Sigma^{(2)}$ and $\hat \Sigma^{(\infty)}$ are not very different.

We also present in Table \ref{BaRaE1} the results of our previous calculations
for E1 transition amplitudes \cite{DzubaRa}. In spite of the very simple 
approximation for the wave functions used in the previous work, the 
agreement for the amplitudes is generally remarkably good. The
exception is the amplitudes which involve a change of spin. 
These amplitudes are larger in the present calculations than in our 
previous work. The reason is the underestimation of relativistic
effects for the $5d$ state of Ba and $6d$ state of Ra in Ref. \cite{DzubaRa}. 
The electric dipole transitions between states
of different spin are forbidden in the non-relativistic limit.
Therefore, larger amplitudes means larger relativistic effects.
Since we don't have experimental values for the amplitudes, 
we can use fine structure intervals instead to see how well relativistic
effects are treated in different calculations. One can see from
the data given in Table I of Ref. \cite{DzubaRa} that the fine structure 
intervals between the ${^3D}_{1,2,3}$ states of Ba and Ra are about two
times smaller than the experimental values. In contrast, all fine structure
intervals of the present calculations are very close to experiment
(see Table \ref{bara}). Therefore, we expect the corresponding amplitudes to be 
more accurate.

\begin{table}
\caption{E1-transition amplitudes for Ba and Ra in different approximations
($|\langle i||d_z|| j \rangle| a_0$).}
\label{BaRaE1}
\begin{ruledtabular}
\begin{tabular}{cccccc}
\multicolumn{2}{c}{Transition} & CI &  
$\hat \Sigma^{(2)}$ & $\hat \Sigma^{(\infty)}$ & Other~\cite{DzubaRa} \\
\hline
\multicolumn{6}{c}{Barium} \\
 $^3$P$_0$ & $^3$D$_1$ &  2.6185  &  2.3149  &  2.3045  &  2.3121 \\
 $^3$P$_1$ & $^1$S$_0$ &  0.3203  &  0.5281  &  0.5240  &  0.4537 \\

 $^3$P$_1$ & $^3$D$_1$ &  2.2829  &  2.0104  &  2.0026  &  2.0108 \\
 $^3$P$_1$ & $^3$D$_2$ &  3.8806  &  3.4309  &  3.4128  &  3.4425 \\

 $^3$P$_1$ & $^1$D$_2$ &  0.2979  &  0.4675  &  0.4999  &  0.1610 \\

 $^3$P$_2$ & $^3$D$_1$ &  0.5997  &  0.5262  &  0.5247  &  0.5275 \\
 $^3$P$_2$ & $^3$D$_2$ &  2.2838  &  2.0012  &  1.9933  &  2.024  \\
 $^3$P$_2$ & $^3$D$_3$ &  5.4285  &  4.8181  &  4.7805  &  4.777  \\

 $^3$P$_2$ & $^1$D$_2$ &  0.3321  &  0.3551  &  0.3402  &  0.1573 \\

 $^1$P$_1$ & $^1$S$_0$ &  5.7133  &  5.4235  &  5.4695  &  5.236  \\
 $^1$P$_1$ & $^3$D$_1$ &  0.0880  &  0.0850  &  0.0735  &  0.1047 \\
 $^1$P$_1$ & $^3$D$_2$ &  0.5935  &  0.4143  &  0.3992  &  0.4827 \\
 $^1$P$_1$ & $^1$D$_2$ &  0.9919  &  1.3062  &  1.1394  &  1.047  \\
\multicolumn{6}{c}{Radium} \\
 $^3$P$_0$ & $^3$D$_1$ &  3.2996  &  2.9325  &  2.9521  &  3.0449 \\
 $^3$P$_1$ & $^1$S$_0$ &  0.8241  &  1.2317  &  1.2205  &  1.0337 \\
 $^3$P$_1$ & $^3$D$_1$ &  2.8836  &  2.5155  &  2.5366  &  2.6389 \\
 $^3$P$_1$ & $^3$D$_2$ &  4.8393  &  4.2931  &  4.3158  &  4.4399 \\
 $^3$P$_1$ & $^1$D$_2$ &  0.7095  &  0.7397  &  0.8068  &  0.0467 \\
 $^3$P$_2$ & $^3$D$_1$ &  0.7799  &  0.6714  &  0.6781  &  0.7166 \\
 $^3$P$_2$ & $^3$D$_2$ &  2.9438  &  2.5357  &  2.5615  &  2.7283 \\
 $^3$P$_2$ & $^3$D$_3$ &  6.9465  &  6.2626  &  6.2541  &  6.3728 \\
 $^3$P$_2$ & $^1$D$_2$ &  0.4285  &  0.5885  &  0.5344  &  0.1499 \\
 $^1$P$_1$ & $^1$S$_0$ &  5.7703  &  5.3652  &  5.4821  &  5.4797 \\
 $^1$P$_1$ & $^3$D$_1$ &  0.3736  &  0.4381  &  0.4255  &  0.4441 \\
 $^1$P$_1$ & $^3$D$_2$ &  0.6162  &  0.3965  &  0.3591  &  1.188  \\
 $^1$P$_1$ & $^1$D$_2$ &  2.9923  &  3.3103  &  3.1379  &  2.4053 \\
\end{tabular}
\end{ruledtabular}
\end{table}

\subsection{Lifetimes of low-lying states of barium and radium}

\begin{table*}
\caption{Lifetimes of low-lying states of barium. Numbers in square brackets denote powers of 10.}
\label{tBa}
\begin{ruledtabular}
\begin{tabular}{cccccccccc}
State & \multicolumn{6}{c}{Lower states to decay to} & 
\multicolumn{3}{c}{Lifetime} \\
      & State & Transition & $\omega$ & Amplitude 
& \multicolumn{2}{c}{Probability (s$^{-1}$)}  & This work &
Dzuba {\em et al.}~\cite{DzubaRa} & Other \\
     &       &            & (a.u.) & (a.u.) & This work & Ref.~\cite{Bizzari} & & & \\
\hline
$^3$P$_0$ & $^3$D$_1$ & E1 & 0.01454 & 2.305  & 3.500[ 5] & & 2.86~$\mu$s & 2.83~$\mu$s & \\

$^3$P$_1$ & $^1$S$_0$ & E1 & 0.05758 & 0.5241 & 3.743[ 5] & & 1.23~$\mu$s & 1.37~$\mu$s & 
1.2~$\mu$s\footnotemark[1] \\
        & $^3$D$_1$ & E1 & 0.01642 & 2.003  & 1.267[ 5] & &    & & \\
        & $^3$D$_2$ & E1 & 0.01559 & 3.413  & 3.150[ 5] & &    & & \\
        & $^1$D$_2$ & E1 & 0.00567 & 0.4999 & 3.257[ 2] & &    & & \\

$^3$P$_2$ & $^3$D$_1$ & E1 & 0.02041 & 0.5247 & 1.003[ 4] & & 1.44~$\mu$s & 1.41~$\mu$s & \\
          & $^3$D$_2$ & E1 & 0.01958 & 1.993  & 1.278[ 5] & &    & & \\
          & $^3$D$_3$ & E1 & 0.01785 & 4.781  & 5.570[ 5] & &    & & \\
          & $^1$D$_2$ & E1 & 0.00967 & 0.3403 & 4.484[ 2] & &    & & \\

$^1$P$_1$ & $^1$S$_0$ & E1 & 0.08238 & 5.470  & 1.194[ 8] & 1.19(1)[8] & 8.35~ns & 9.1~ns & 
8.37(8)~ns\footnotemark[2] \\
        & $^3$D$_1$ & E1 & 0.04122 & 0.0735 & 2.702[ 3] & 3.1(1)[3] &     & & \\
        & $^3$D$_2$ & E1 & 0.04039 & 0.3993 & 7.498[ 4] & 1.1(2)[5] &     & & \\
        & $^1$D$_2$ & E1 & 0.03047 & 1.139  & 2.623[ 5] & 2.5(2)[5] &     & & \\

$^3$D$_2$ & $^1$S$_0$ & E2 & 0.04199 & 3.125  & 1.454[-2] & & 69~s & & \\

\end{tabular}
\end{ruledtabular}
\noindent \footnotetext[1]{Reference \cite{Radzig}.}
\noindent \footnotetext[2]{Reference \cite{Niggli}.}
\end{table*}
\begin{table*}
\caption{Lifetimes of low-lying states of radium. Numbers in square brackets denote powers of 10.}
\label{tRa}
\begin{ruledtabular}
\begin{tabular}{ccccccccc}
State & \multicolumn{5}{c}{Lower states to decay to} & 
\multicolumn{3}{c}{Lifetime} \\
      & State & Transition & $\omega$ & Amplitude 
& Probability  & This work &
Dzuba {\em et al.}~\cite{DzubaRa} & Other \\
     &       &            & (a.u.) & (a.u.) & (s$^{-1}$) & & & \\
\hline
$^3$D$_1$ & $^3$P$_0$ & E1 & 0.00291 & 2.952  & 1.529[ 3] & 654~$\mu$s & 617~$\mu$s & \\

$^1$D$_2$ & $^3$P$_1$ & E1 & 0.01404 & 0.8068 & 7.722[ 3] & 129~ms & 38~ms & \\
          & $^3$P$_2$ & E1 & 0.00179 & 0.5345 & 7.973[ 0] &      & & \\

$^3$P$_1$ & $^1$S$_0$ & E1 & 0.06378 & 1.221  & 2.760[ 6] & 362~ns & 505~ns &
420~ns\footnotemark[1],250~ns\footnotemark[2] \\
          & $^3$D$_1$ & E1 & 0.00129 & 2.537  & 9.850[ 1] &      & & \\
          & $^3$D$_2$ & E1 & 0.00002 & 4.316  & 1.572[-3] &      & & \\

$^3$P$_2$ & $^3$D$_1$ & E1 & 0.01355 & 0.6782 & 4.897[ 3] & 5.55~$\mu$s & 5.2~$\mu$s & \\
          & $^3$D$_2$ & E1 & 0.01228 & 2.562  & 5.204[ 4] &      & & \\
          & $^3$D$_3$ & E1 & 0.00903 & 6.254  & 1.234[ 5] &      & & \\

$^1$P$_1$ & $^1$S$_0$ & E1 & 0.09439 & 5.482  & 1.805[ 8] & 5.53~ns & 5.5~ns & \\
          & $^3$D$_1$ & E1 & 0.03189 & 0.4256 & 4.195[ 4] &      & & \\
          & $^3$D$_2$ & E1 & 0.03063 & 0.3592 & 2.646[ 4] &      & & \\
          & $^1$D$_2$ & E1 & 0.01656 & 3.138  & 3.194[ 5] &      & & \\

$^3$D$_2$ & $^1$S$_0$ & E2 & 0.06376 & 5.022  & 3.032[-1] & 3.3~s & 15~s &
4~s\footnotemark[3] \\

\end{tabular}
\end{ruledtabular}
\noindent \footnotetext[1]{Reference \cite{Hafner}.}
\noindent \footnotetext[2]{Reference \cite{Bruneau}.}
\noindent \footnotetext[3]{Reference \cite{F-F}.}
\end{table*}

The lifetime of the atomic state $i$ expressed in seconds is given by
\begin{equation}
  \tau_i =  2.4189 \times 10^{-17}/{\sum_j T_{ij}},
\label{tau}
\end{equation}
where $T_{ij}$ is the probability of a transition from state $i$ to
a lower state $j$ (in atomic units), the numerical factor is to convert 
atomic units to seconds, and summation goes over all states $j$
that have energies lower than the energy of state $i$.

In the present paper we consider only electric dipole (E1) and electric
quadrupole (E2) transitions.
The probability of the E1 transition from state $i$ to a lower state $j$
is (atomic units)
\begin{equation}
  T_{ij} = \frac{4}{3} (\alpha \omega_{ij})^3 \frac{A_{ij}^2}{2J_i+1},
\label{E1}
\end{equation}
where $\omega_{ij} = \epsilon_i - \epsilon_j$, $A_{ij}$ is the amplitude
of the transition (reduced matrix element of the electric dipole operator),
and $J_i$ is the value of the total angular momentum of the state $i$.
The probability of the E2 transition is (atomic units)
\begin{equation}
  T_{ij} = \frac{1}{15} (\alpha \omega_{ij})^5 \frac{A_{ij}^2}{2J_i+1}.
\label{E2}
\end{equation}

Lifetimes of low-lying states of barium and radium calculated using
transition amplitudes from Table \ref{BaRaE1} and experimental
energies are presented in Tables \ref{tBa} and \ref{tRa}.
The new data show systematic improvement of the agreement between
theory and experiment compared to our previous work \cite{DzubaRa}.
However, the change is small. The only significant change is for
the $^3P_1$ states of barium and radium and $^3D_2$ state of
radium. This is due to the change in the E1 amplitude of the 
$^3P_1 \ - \ ^1S_0$ transition and the E2 amplitude of the
$^3D_2 \ - \ ^1S_0$ transition. The new values
are more accurate due to the better treatment of relativistic effects 
(see discussion above).

\section{Acknowledgments}

We are grateful to V. Flambaum and R. Holt for stimulating discussions.
This work is supported by the Australian Research Council. 
J.G. acknowledges support from an Avadh Bhatia Women's Fellowship and from 
Science and Engineering Research Canada while at University of Alberta.



\begin{thebibliography}{20}

\bibitem{Ginges} J.S.M. Ginges and V.V. Flambaum, Phys. Rep. \textbf{397}, 63 (2004).

\bibitem{FlambaumRa} V.V. Flambaum, Phys. Rev. A \textbf{60}, R2611 (1999).

\bibitem{DzubaRa} V.A. Dzuba, V.V. Flambaum, and J.S.M. Ginges,
Phys. Rev. A \textbf{61}, 062509 (2000).

\bibitem{argonne} I. Ahmad, K. Bailey, J.R. Guest, R.J. Holt, Z.-T. Lu, T. O'Connor, 
D. Potterveld, E.C. Schulte, and N.D. Scielzo, 
http://www-mep.phy.anl.gov/atta/research/radiumedm.html.

\bibitem{groningen} K. Jungmann, G.P. Berg, U. Damalapati, P. Dendooven, O. Dermois, 
M.N. Harakeh, R. Hoekstra, R. Morgenstern, A. Rogachevskiy, M. Sanchez-Vega, R. Timmermans, E. Traykov, 
L. Willmann, and H. W. Wilschut, Phys. Scripta {\bf T104}, 178 (2003).

\bibitem{Moore} C. E. Moore, {\em Atomic Energy Levels}, Natl. Bur.
Stand. (U.S.), Circ. No. 467 (U.S. GPO, Washington, D. C., 1958),
Vols. 1-3. 

\bibitem{Rasmussen} E. Rusmussen, Zeit. Phys. {\bf 87}, 607 (1934).

\bibitem{Russell} H.N. Russell, Phys. Rev. {\bf 46}, 989 (1934).

\bibitem{F-F} J. Biero\'{n}, C. Froese Fischer, S. Fritzsche, and K. Pachucki,
J. Phys. B \textbf{37}, L305 (2004).

\bibitem{Johnson98} V.A. Dzuba, and W.R. Johnson,
Phys. Rev. A {\textbf{57}}, 2459 (1998).

\bibitem{vn} V.A. Dzuba, Phys. Rev. A {\bf 71} 032512 (2005).

\bibitem{Kozlov96} V.A. Dzuba, V.V. Flambaum, and M.G. Kozlov,
Phys. Rev. A {\textbf{54}}, 3948 (1996).

\bibitem{CPM} V.A. Dzuba, V.V. Flambaum, P.G. Silvestrov, and 
O.P. Sushkov, J. Phys. B {\textbf{20}}, 3297 (1987).

\bibitem{Dzuba89} V.A. Dzuba, V.V. Flambaum, and O.P. Sushkov, 
Phys. Lett. A {\textbf{140}}, 493 (1989).

\bibitem{DzubaFr} V.A. Dzuba, V.V. Flambaum, and O.P. Sushkov,
Phys. Rev. A \textbf{51}, 3454 (1995).

\bibitem{DzubaCs} V.A. Dzuba, V.V. Flambaum, and J.S.M. Ginges,
Phys. Rev. D \textbf{66}, 076013 (2002).

\bibitem{KozlovRa} V.A. Dzuba, V.V. Flambaum, J.S.M. Ginges,
and M.G. Kozlov, Phys. Rev. A \textbf{66}, 012111 (2002).

\bibitem{Bizzari} A. Bizzari and M.C.E. Huber, Phys. Rev. A \textbf{42}, 5422 (1990).

\bibitem{Radzig} A.A. Radzig and B.M. Smirnov, 
Reference Data on Atoms, Molecules and Ions (Springer, Berlin, 1985).

\bibitem{Niggli} S. Niggli, and M.C.E. Huber, Phys. Rev. A \textbf{39}, 3924 (1989).

\bibitem{Hafner} P. Hafner and W.H.E. Schwarz, J. Phys. B {\bf 11}, 2975 (1978).

\bibitem{Bruneau} J. Bruneau, J. Phys. B {\bf 17}, 3009 (1984).
\end{thebibliography}
\end{document}